\documentclass[a4paper,11pt]{article}

\usepackage{contribution}
\graphicspath{{plots/}}

\newcommand{\weblink}[2][]{%
    \ifthenelse{\equal{#1}{}}%
    {\textnormal{\url{#2}}}%
    {\textnormal{\href{#2}{#1}}}%
}

\newcommand{\acknowledgements}[1]{%
  \bigskip\bigskip
  \textsf{\textbf{\Large Acknowledgements}} \\[2ex]
  {#1}
  \bigskip
}

\def\beq{\begin{equation}}
\def\eeq#1{\label{#1}\end{equation}}
\def\eeqn{\end{equation}}

\def\beqa{\begin{eqnarray}}
\def\eeqa#1{\label{#1}\end{eqnarray}}
\def\eeqan{\end{eqnarray}}

\let\bar=\overbar

\def\Dslash{\not{\hbox{\kern-4pt $D$}}}
\def\dslash{\not{\hbox{\kern-2pt $\del$}}}

\def\msb{{\bar{\ssstyle M \kern -1pt S}}}

\newcommand{\contribution}[7][]{%
  \clearpage
  \thispagestyle{plain}
  \ifthenelse{\equal{#1}{}}
  {\hypersetup{pdftitle={#2}}}
  {\hypersetup{pdftitle={#1}}}
  \hypersetup{pdfauthor={{#3} {#4}}}
  {\centering\normalfont\LARGE\bfseries\sffamily #2 \par\nobreak}
  \lhead{}
  \chead{%
    \textit{\footnotesize XIV International Conference on Hadron Spectroscopy
      (\weblink[\textit{hadron2011}]{http://www.hadron2011.de}), 13-17 June 2011, Munich, Germany}%
  }
  \rhead{}
  \bigskip
  \begin{center}
    {#3} {#4}\ifthenelse{\equal{#6}{}}{}{\footnote{\weblink[#6]{mailto:#6}}}
    \ifthenelse{\equal{#7}{}}{}{#7} \\
    \textit{#5}
  \end{center}
  \bigskip
}

\renewcommand{\abstract}[1]{%
  \begin{center}
    \begin{minipage}{0.85\textwidth}
      \begin{footnotesize}
        #1
      \end{footnotesize}
    \end{minipage}
  \end{center}
  \bigskip
}

\begin{document}

% % % % % % % % % % % % % % % % % % % % % % % % % % % % % % % % % % % % % % % % %
% your proceedings
%%%%%%%%%%%%%%%%%%%%%%%%%%%%%%%%%%%%%%%%%%%%%%%%%%%%%%%%%%%%%%%%%%%%%%%%%%%%%%%%%
%
% template for hadron2011 contribution
%
% please do not rename this file
%
% to create document run
%
%     pdflatex hadron2011.tex
%
%%%%%%%%%%%%%%%%%%%%%%%%%%%%%%%%%%%%%%%%%%%%%%%%%%%%%%%%%%%%%%%%%%%%%%%%%%%%%%%%%
{  % do not remove

%%%%%%%%%%%%%%%%%%%%%%%%%%%%%%%%%%%%%%%%%%%%%%%%%%%%%%%%%%%%%%%%%%%%%%%%%%%%%%%%%
% please define your macros here

%
%%%%%%%%%%%%%%%%%%%%%%%%%%%%%%%%%%%%%%%%%%%%%%%%%%%%%%%%%%%%%%%%%%%%%%%%%%%%%%%%%

%%%%%%%%%%%%%%%%%%%%%%%%%%%%%%%%%%%%%%%%%%%%%%%%%%%%%%%%%%%%%%%%%%%%%%%%%%%%%%%%%
% define title, author, and address
% contribution[short title]{title}{author first name}{author last name}{author address}{author email}{collaboration}
% the short title will appear in the page headers and the TOC of the book of proceedings
% the last two arguments may be left empty
\contribution[Diffractive Dissociation into 3 Pion Final States at COMPASS]  % short title (optional)
{Diffractive Dissociation into $\mathbold{\pi^- \pi^- \pi^+}$ Final State\\ at COMPASS}  % title
{Florian}{Haas}  % first and last name of author
{Physik Department E18 \\
  Technische Universität München \\
  D-85748 Garching, GERMANY}  % author address
{florian.haas@tum.de}  % author email optional
{on behalf of the COMPASS Collaboration}  % collaboration (optional)
%
%%%%%%%%%%%%%%%%%%%%%%%%%%%%%%%%%%%%%%%%%%%%%%%%%%%%%%%%%%%%%%%%%%%%%%%%%%%%%%%%%

%%%%%%%%%%%%%%%%%%%%%%%%%%%%%%%%%%%%%%%%%%%%%%%%%%%%%%%%%%%%%%%%%%%%%%%%%%%%%%%%%
% abstract
\abstract{%
Diffractive dissociation reactions studied at the COMPASS experiment at CERN provide access to the light-meson spectrum.
During a pilot run in 2004, using a negative pion beam and a Pb target, 420k $\pi^- \pi^- \pi^+$ final-state events with masses below 2.5 GeV/$c^2$ were recorded,
yielding a significant spin-exotic signal for the controversial $\pi_1$(1600) resonance. After a major upgrade of the spectrometer in 2007, 
the following two years were dedicated to hadron spectroscopy. Using again a pion beam, but now with a liquid hydrogen target, 
a unique statistics of $\sim$60M events of the 3$\pi$ final state was gathered in 2008. During a short campaign in 2009, 
the H$_2$ target was replaced by several solid state targets in order to study the effect of the target material on the production. 
A partial-wave analysis (PWA) was performed on all these data sets and results are presented.
}
%
%%%%%%%%%%%%%%%%%%%%%%%%%%%%%%%%%%%%%%%%%%%%%%%%%%%%%%%%%%%%%%%%%%%%%%%%%%%%%%%%%

%%%%%%%%%%%%%%%%%%%%%%%%%%%%%%%%%%%%%%%%%%%%%%%%%%%%%%%%%%%%%%%%%%%%%%%%%%%%%%%%%
% main text
% for short contributions sections are optional
\section{Introduction}

In the quark model mesons are described as bound states of quarks and anti-quarks. They are characterized by a set of quantum numbers, which are isospin $I$, $G$-parity (for light unflavoured mesons), the total spin $J$, parity $P$ and, for neutral mesons, charge conjugation parity $C$. 
Quantum Chromo Dynamics (QCD) predicts the existence of states beyond the quark model, like hybrids, i.e. systems consisting of a color octet $q \overline q$ pair neutralized in color by gluonic excitation, or glue balls, consisting only of glue.
The experimental identification of such states, however, is difficult due to mixing with $q \overline q$ configurations with the same quantum numbers.
The observation of meson states whose quantum numbers can not be explained within the quark model, e.g. $J^{PC} =$ $0^{--}, 0^{+-}$, $1^{-+}$, ..., would be an evidence for quark-gluon configurations beyond the quark model.\\
The lightest hybrid, predicted in a mass region of 1.3 - 2.2 GeV/$c^2$, is expected to have exotic quantum numbers $J^{PC} = 1^{-+}$ \cite{hybrid}, and therefore does not mix with pure $q \overline q$ states. There are three experimental candidates for a light $1^{-+}$ hybrid. The $\pi_1(1400)$ was observed by E852 \cite{3}, VES \cite{4}, and Crystal Barrel \cite{5}. Another $1^{-+}$ state, the $\pi_1(1600)$, decaying into $\rho \pi$ \cite{7,8,9}, $\eta ' \pi$ \cite{10,11}, $f_1(1285) \pi$ \cite{12,13} and $b_1(1235) \pi$ \cite{13} was observed in peripheral $\pi^- p$ interactions in E852 and VES. The resonant nature of both states, however, is still heavily disputed in the community \cite{4,13,15}. A third exotic state, $\pi_1(2000)$, decaying to $f_1 \pi$ and $b_1 \pi$ was seen in only one experiment \cite{12}.
The COMPASS experiment will contribute significantly to the search for exotic mesons and glue balls in the light-meson sector.\\
COMPASS, the \textbf{CO}mmon \textbf{M}uon and \textbf{P}roton \textbf{A}pparatus for \textbf{S}tructure and \textbf{S}pectroscopy \cite{compass} is located at the CERN SPS accelerator. It is a two-stage magnetic spectrometer which provides a large angular acceptance over a wide momentum range. In addition to calorimetry and particle identification, COMPASS is equipped with a very precise charged particle tracking system. For the beam tracking, Silicon detectors around the target, Scintillating Fibers and PixelGEMs are used. Close to the beam large-size GEM detectors and Micromegas are the backbone of the tracking system. The periphery is covered by Drift Chambers and MWPCs.

\section{The $\mathbold{\pi^- \pi^- \pi^+}$ Final State}

Diffractive dissociation is a process where an incoming beam
particle impinges on a target and is excited to an intermediate state $X$ which finally decays 
into an $n$-body final state. Here we focus on the diffraction of a 190 GeV/$c$ $\pi^-$ beam into the $\pi^- \pi^- \pi^+$ final state. The target particle stays intact, taking away the recoil momentum. In Fig.~\ref{t_3pic_2004} the distribution of the kinematic variable $t'$ is shown, which is given by $|t| - |t|_{\text{min}}$ where $t$ is the squared four-momentum transfer to the target nucleus and $|t|_{\text{min}}$ the minimum value of $|t|$ allowed by kinematics for a given mass $m_{X}$. This analysis focuses on events in the ``high-$t'$'' region between 0.1 and 1.0 GeV$^2$/$c^2$ (see Fig.~\ref{t_3pic_2004}) where
E852\cite{8} observed the production of the exotic $\pi_1$(1600) with $J^{PC} = 1^{-+}$. \\ Figure~\ref{mass_3pic_2004} shows the invariant mass spectrum of $\sim 4.2 \cdot 10^5$ events of the 3$\pi$ final state on a Pb target. In addition, the intensity of the isotropic background wave (triangles) is shown, as obtained from a partial-wave analysis, which will be described in the next section. Figure~\ref{mass_3pic_2008} shows the invariant mass distribution for $\sim 60 \cdot 10^6$ events taken with a liquid hydrogen target. Due to an upgrade of the spectrometer it is sensitive to higher masses than during 2004 data taking.
\enlargethispage{1cm} 
\begin{figure}[t!]
\subfloat
[Squared four-momentum transfer $t'$, logarithmic scale.]
{
\label{t_3pic_2004}
\includegraphics[width=0.31\textwidth]{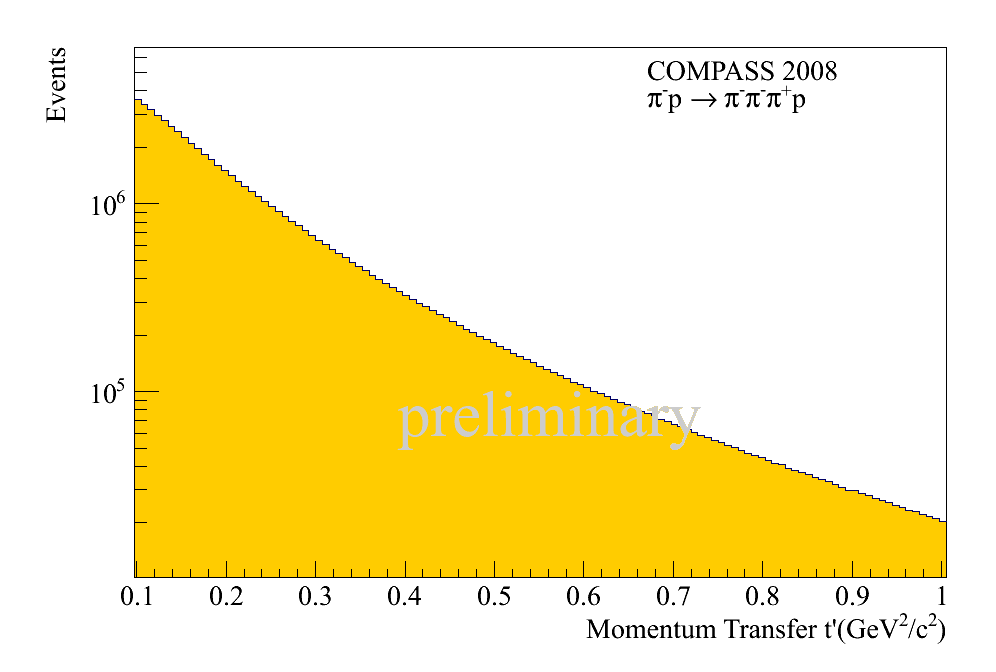}
}
\quad
\subfloat
[$3\pi$ invariant mass spectrum for Pb target (from \cite{3pic}).]
{
\label{mass_3pic_2004}
\includegraphics[width=0.3\textwidth]{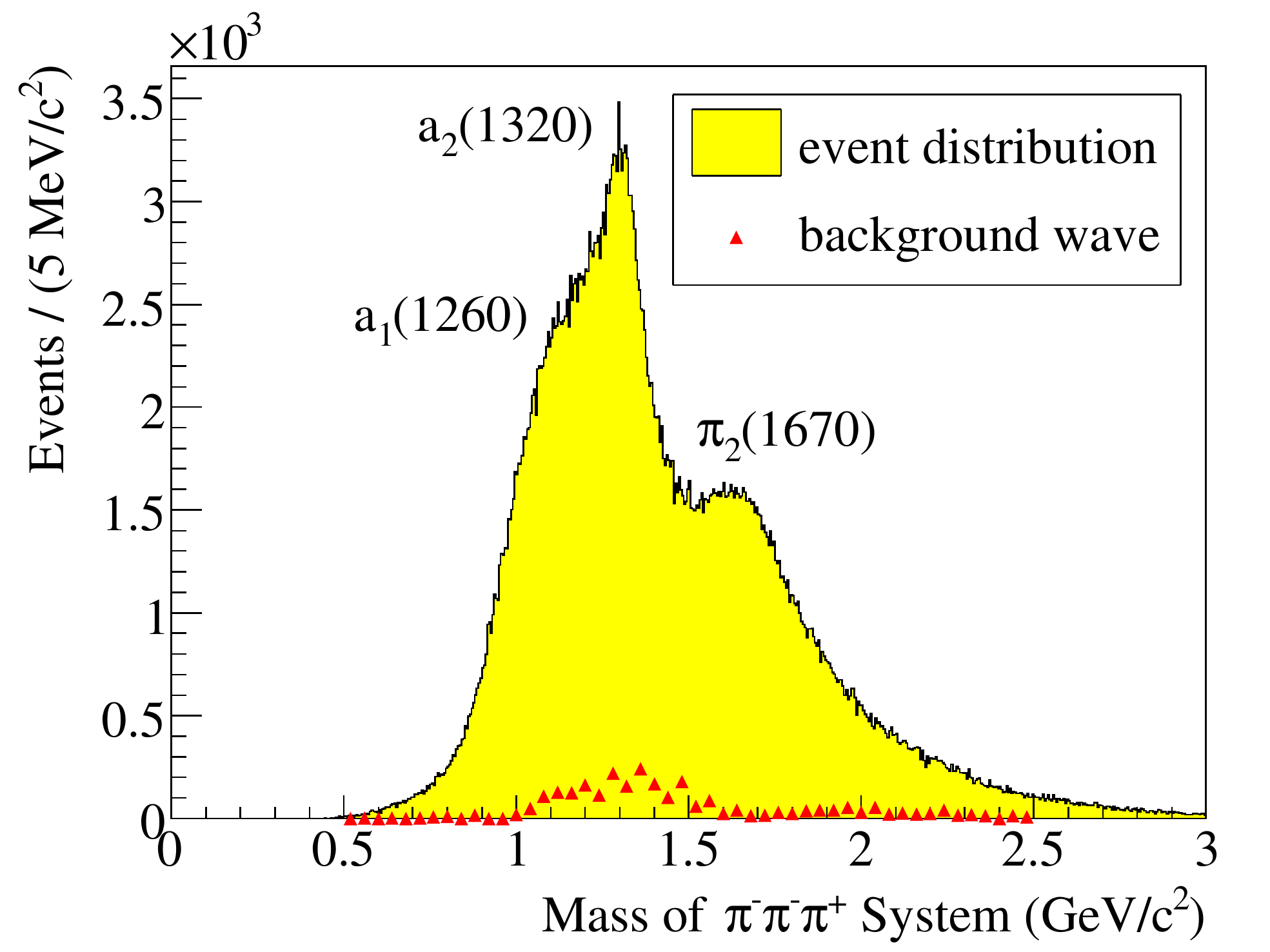}
}
\quad
\subfloat
[$3\pi$ invariant mass spectrum for H$_2$ target.]
{
\label{mass_3pic_2008}
\includegraphics[width=0.31\textwidth]{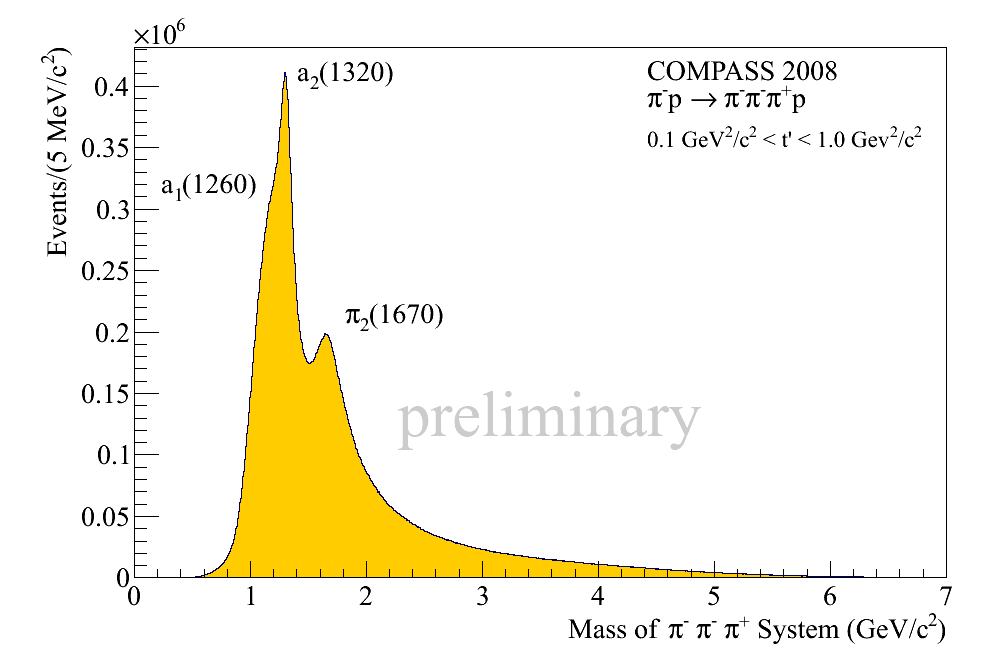}
}
\caption{Kinematic Distributions.}
\end{figure}

\newpage

\subsection{Partial-Wave Technique and Isobar Model}
The present analysis is based on the isobar model, assuming that
the intermediate state $X$ decays first into an isobar and a
bachelor pion, with relative orbital angular momentum $L$ 
between the two. The isobar subsequently decays into a $\pi^- \pi^+$ pair. A partial-wave analysis (PWA), performed in two steps, was applied to the data.
The first step is a fit in bins of the 3$\pi$ invariant mass $m_X$ using the cross section parametrisation:
\begin{equation}
\sigma(\tau,m_X,t')=\sum \limits _{\epsilon=\pm1} \sum \limits _{r=1}^{N_r}\left|\sum \limits _{i} T^{\epsilon}_{ir}(m_X) f_{i}^{\epsilon}(t')\psi_{i}^{\epsilon}(\tau,m) \right|^2 
\label{formula}
\end{equation}
The transition amplitudes $T^{\epsilon}_{ir}$ are the fitting parameters. They are
obtained by an extended maximum likelihood fit. The real functions $f_{i}^{\epsilon}(t')$ describe the $t'$ dependence of the
cross-section. The decay amplitudes $\psi_{i}^{\epsilon}(\tau,m)$ are described by Zemach tensors or $D$ functions. $\tau$ represents the five phase-space variables of the three-body decay. The indices $i$ and $\epsilon$ denote different partial waves, defined by a set of quantum numbers $J^{PC} M^{\epsilon} [$isobar $\pi] L$.\\
 $M$ is the absolute value of the spin projection onto the beam direction; $\epsilon$ is the reflectivity \cite{chung}, which describes the symmetry under reflections through the production plane and corresponds to the naturality of the exchanged particle in the reaction. By incoherent summation over $r$ different spin states of the target are taken into account. \\
In a second step the $m_X$ dependence of the spin-density matrix, as obtained from the first step, is fit using a $\chi^2$ minimization method and a model based on Breit-Wigner functions for the resonant parts plus coherent background terms (mass-dependent fit).

\subsection{Fit Results}
For a data sample obtained with a Pb target during the 2004 campaign, the model, on which the corresponding fit in 40 MeV mass bins is based on, consists of 42 partial waves: 34 waves with natural parity exchange ($\varepsilon=1$) and 7 waves corresponding to unnatural parity exchange. An additional background wave is added incoherently to the cross-section.
Except the latter, the waves are constructed with quantum numbers ranging from total spin $J = 0$ to $J = 4$, spin projection $M = 0$ and $M = 1$ and orbital angular momentum $L = S$ to $L = G$. Five isobars were used, four ($\rho(770), f_0(980), f_2(1270)$ and $\rho_3(1690)$) described by Breit-Wigner functions and in addition a $\pi\pi_S$-wave parametrized with the "K$_1$"-solution \cite{Au} from which the complex amplitude of the $f_0(980)$ was subtracted. Exponential functions specify the $t'$ dependence.
Thresholds on the $3\pi$ mass were applied to most of the waves, in order to stabilize the fit. To the six most prominent waves a mass-dependent fit was applied. Breit-Wigner functions and a coherent background, where necessary, are characterizing the waves.\\
Figures~\ref{fitsum}a-c show the intensity of the three most prominent waves, $1^{++}0^+ \rho \pi S$,\\$2^{++}1^+ \rho \pi D$, and $2^{-+}0^+ f_2 \pi S$.
Of peculiar interest are the fit results for the spin-exotic wave. The $1^{-+}1^+ \rho \pi P$ intensity (Fig.~\ref{pi1}) features a broad bump, centered at
1.7 GeV/$c^2$, with a visible low-mass shoulder. A constant-width Breit-Wigner function (blue
curve) and a non-resonant background (purple curve), possibly caused by Deck-like effects, describe the data well. To clarify the resonant nature of this exotic
amplitude, its interferences with well\\

\begin{figure}[h!]
\subfloat
[]
{
\label{a2}
\includegraphics[width=0.325\textwidth]{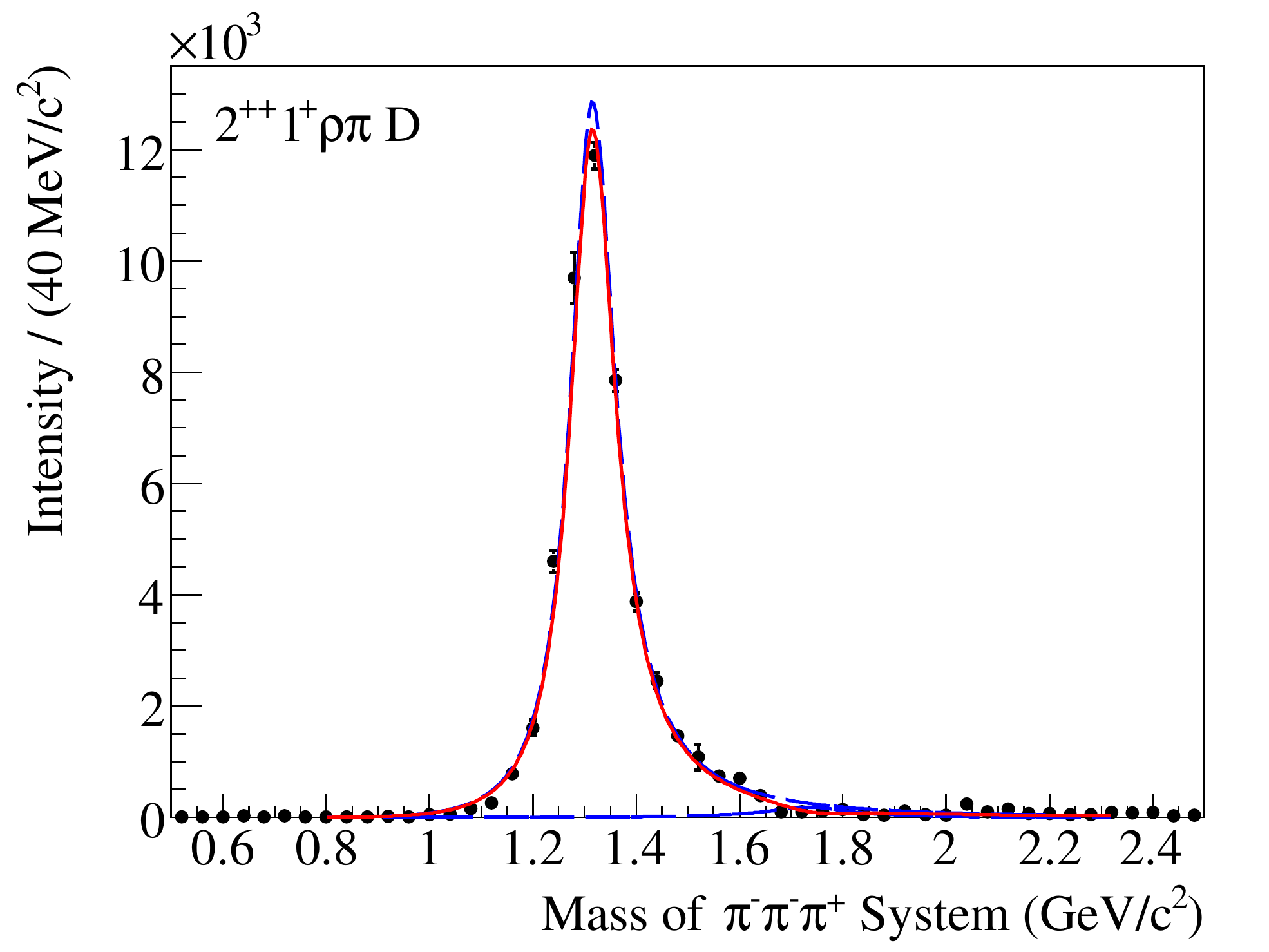}
}
%\qquad
\subfloat
[]
{
\label{a1}
\includegraphics[width=0.325\textwidth]{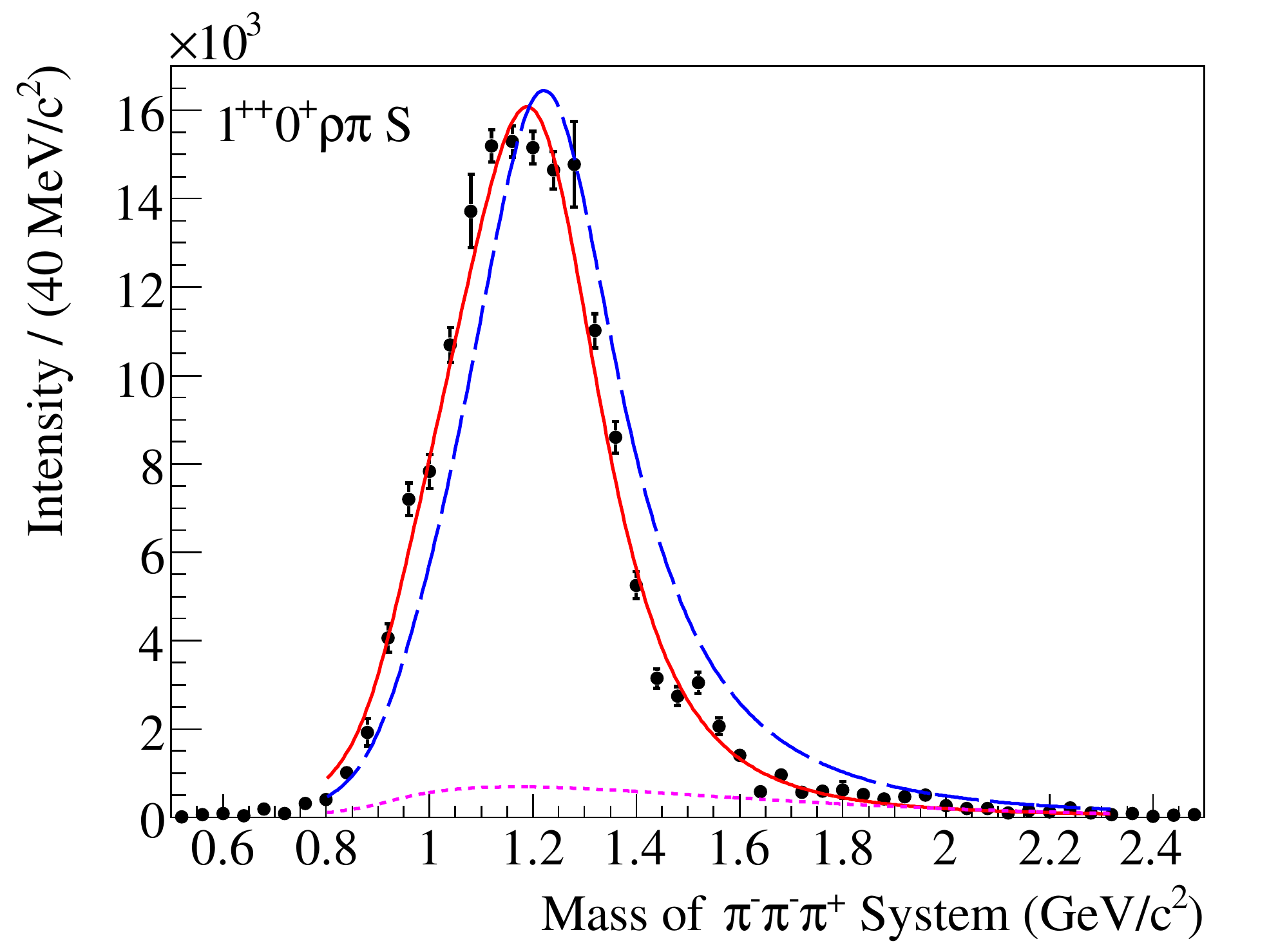}
}
%\qquad
\subfloat
[]
{
\label{pi2}
\includegraphics[width=0.325\textwidth]{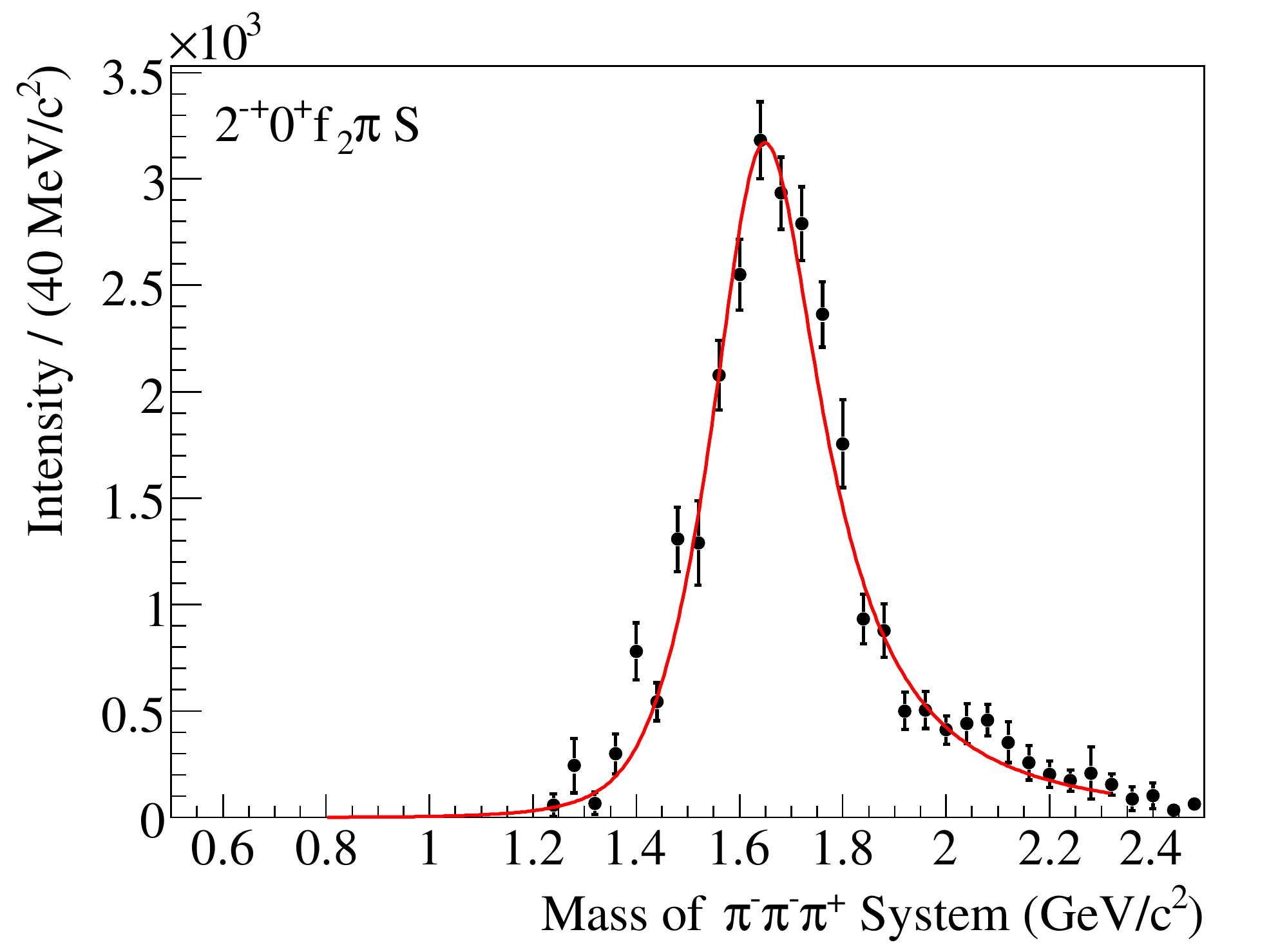}
}

\subfloat
[]
{
\label{pi1}
\includegraphics[width=0.325\textwidth]{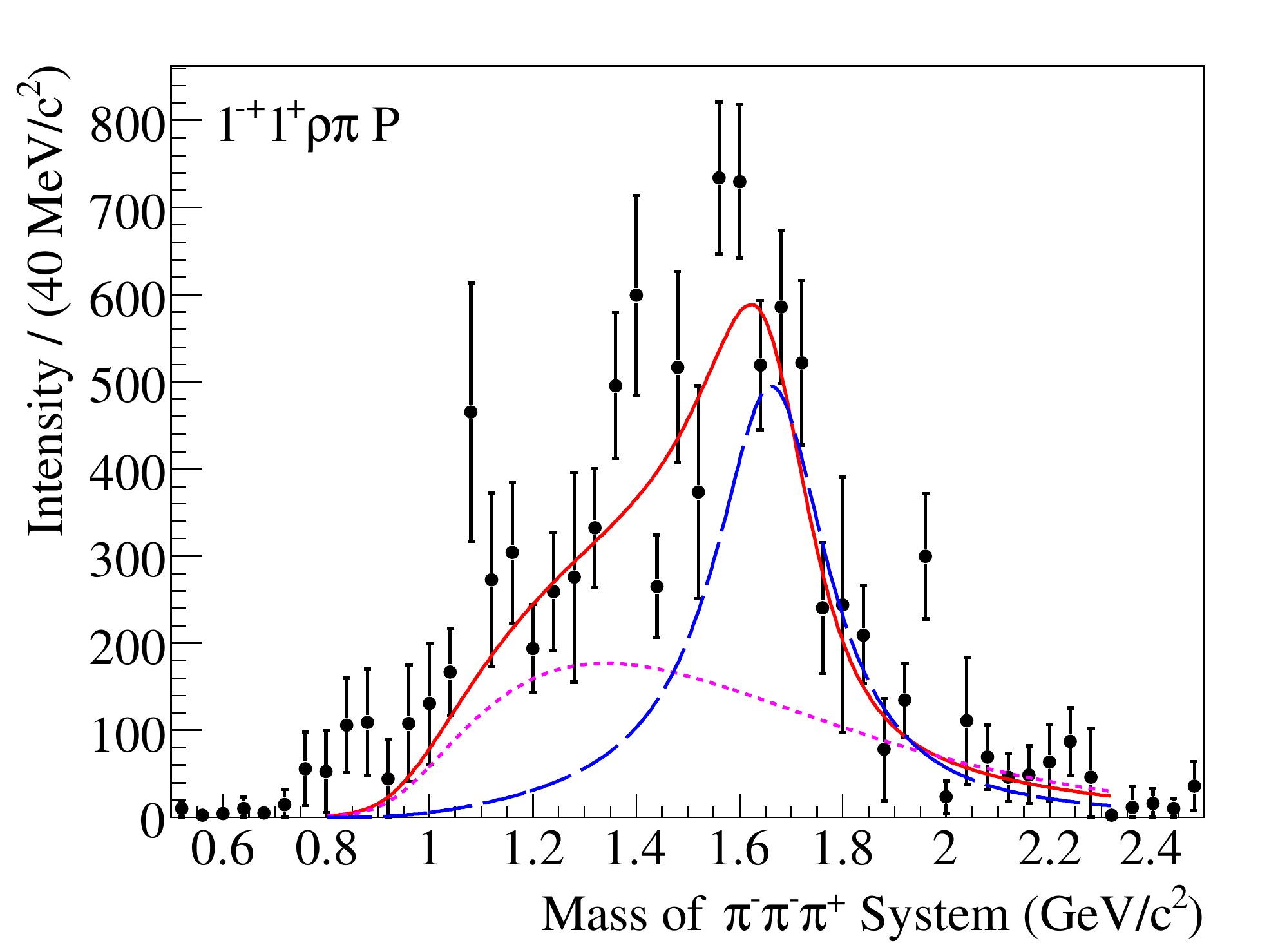}
}
%\qquad
\subfloat
[]
{
\label{phasea1}
\includegraphics[width=0.325\textwidth]{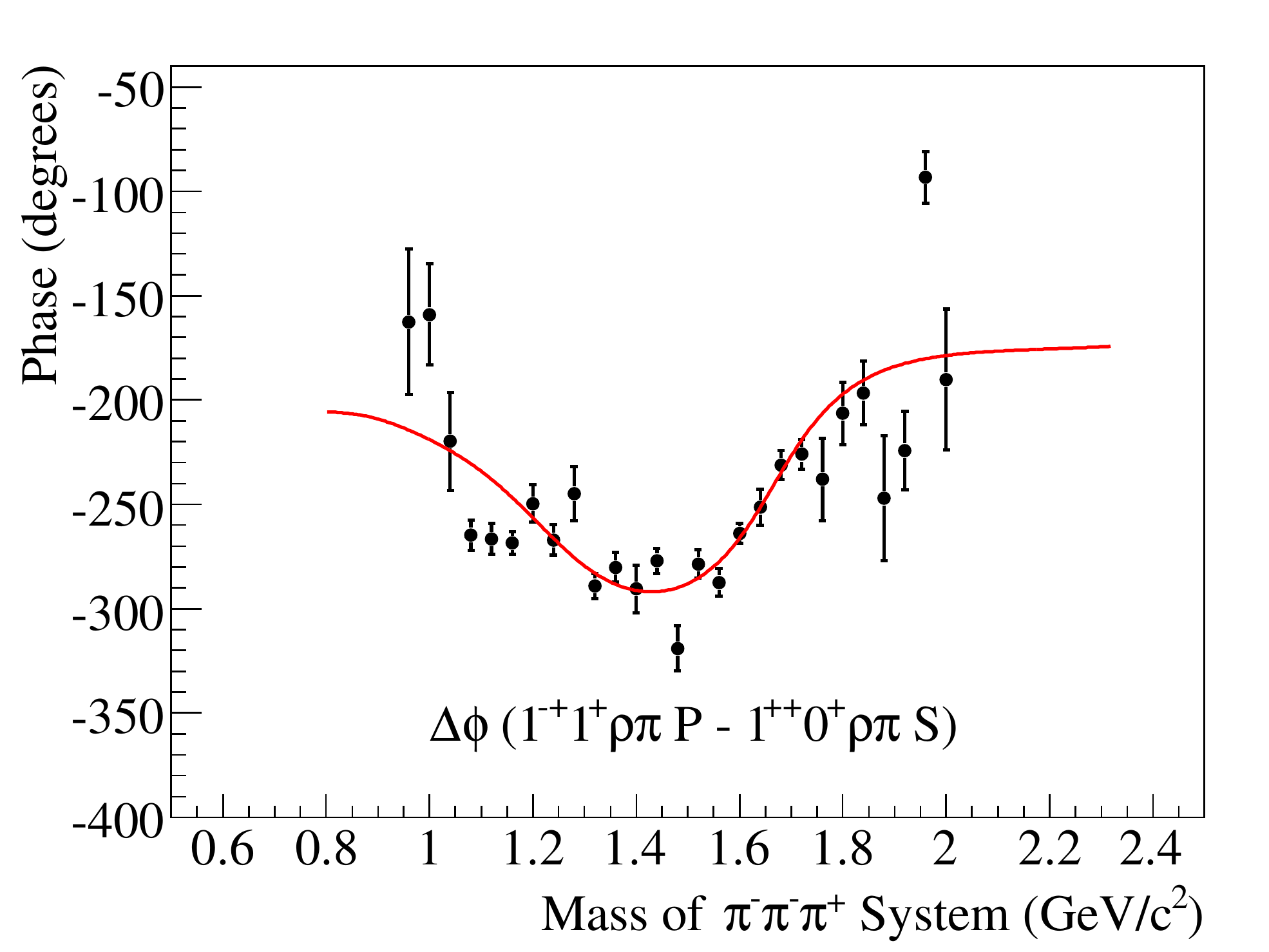}
}
%\qquad
\subfloat
[]
{
\label{phasepi2}
\includegraphics[width=0.325\textwidth]{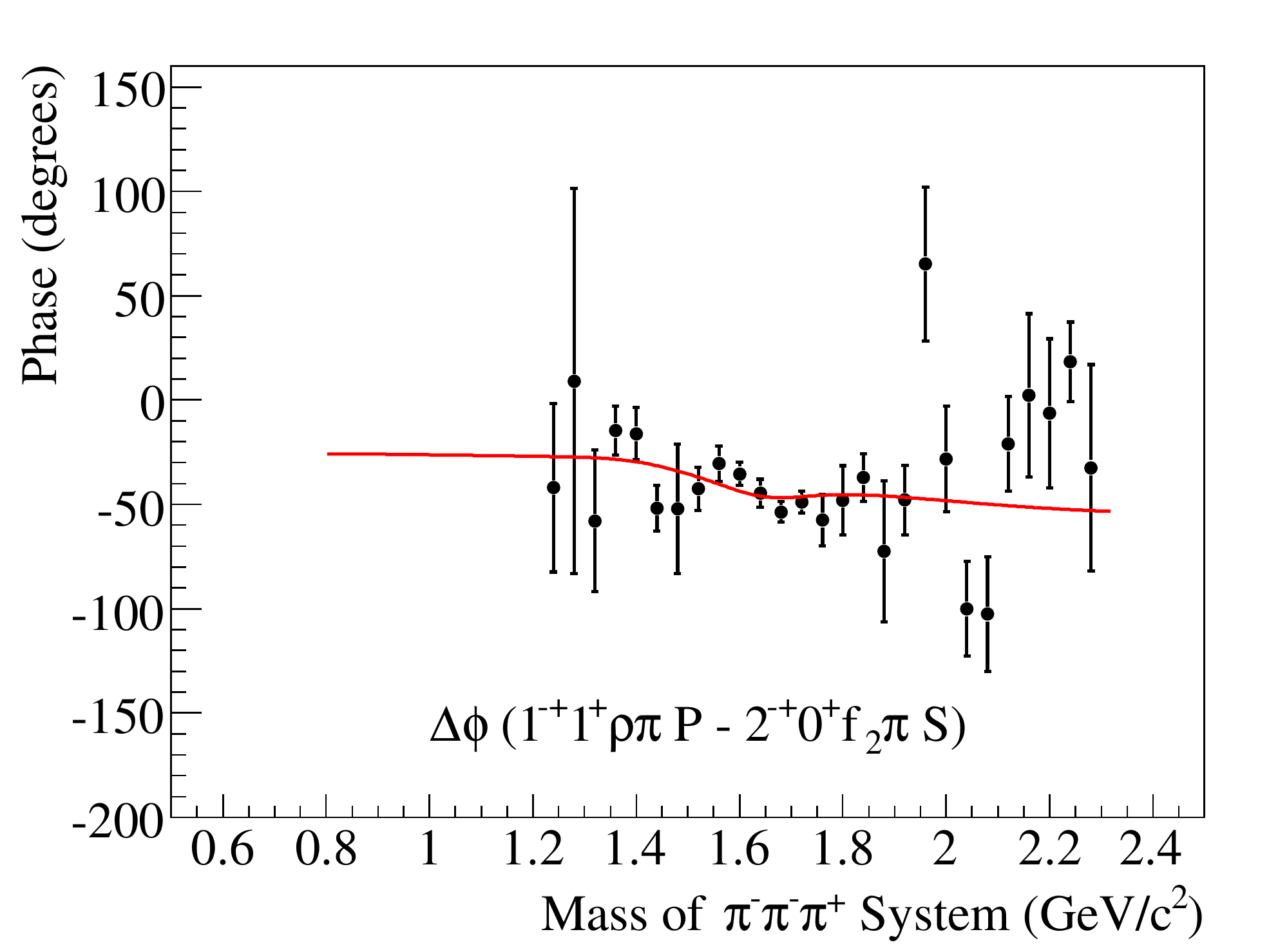}
}
\caption{Intensities of major waves: (a) $2^{++}1^+ \rho \pi D$, (b) $1^{++}0^+ \rho\pi S$ and (c) $2^{-+}0^+ f_2 \pi S$, the red curve represents the mass-dependent fit of the $a_2(1320)$, the $a_1(1260)$ and the $\pi_2(1670)$ respectively. (d) Intensity of the spin-exotic wave $1^{-+}1^+ \rho \pi P$, and phase differences of this wave with respect to the (e) $1^{++}0^+ \rho\pi S$, and the (f) $2^{-+}0^+ f_2 \pi S$ waves (Pb target). All plots from (from \cite{3pic})}
\label{fitsum}
\end{figure}

%*************************2008
\begin{figure}[h!]
\subfloat
[]
{
\label{a2_2008}
\includegraphics[width=0.325\textwidth]{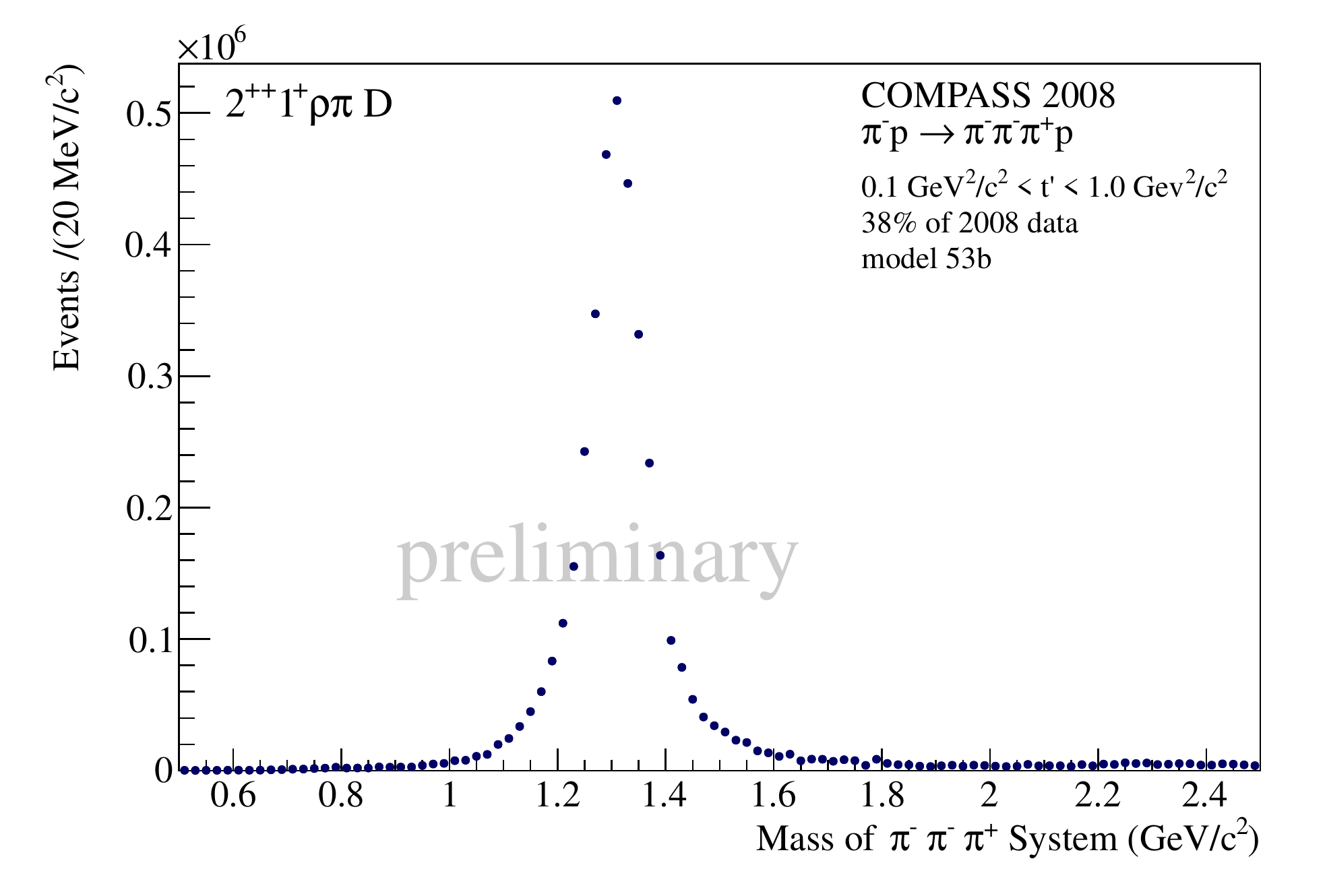}
}
%\qquad
\subfloat
[]
{
\label{a1_2008}
\includegraphics[width=0.325\textwidth]{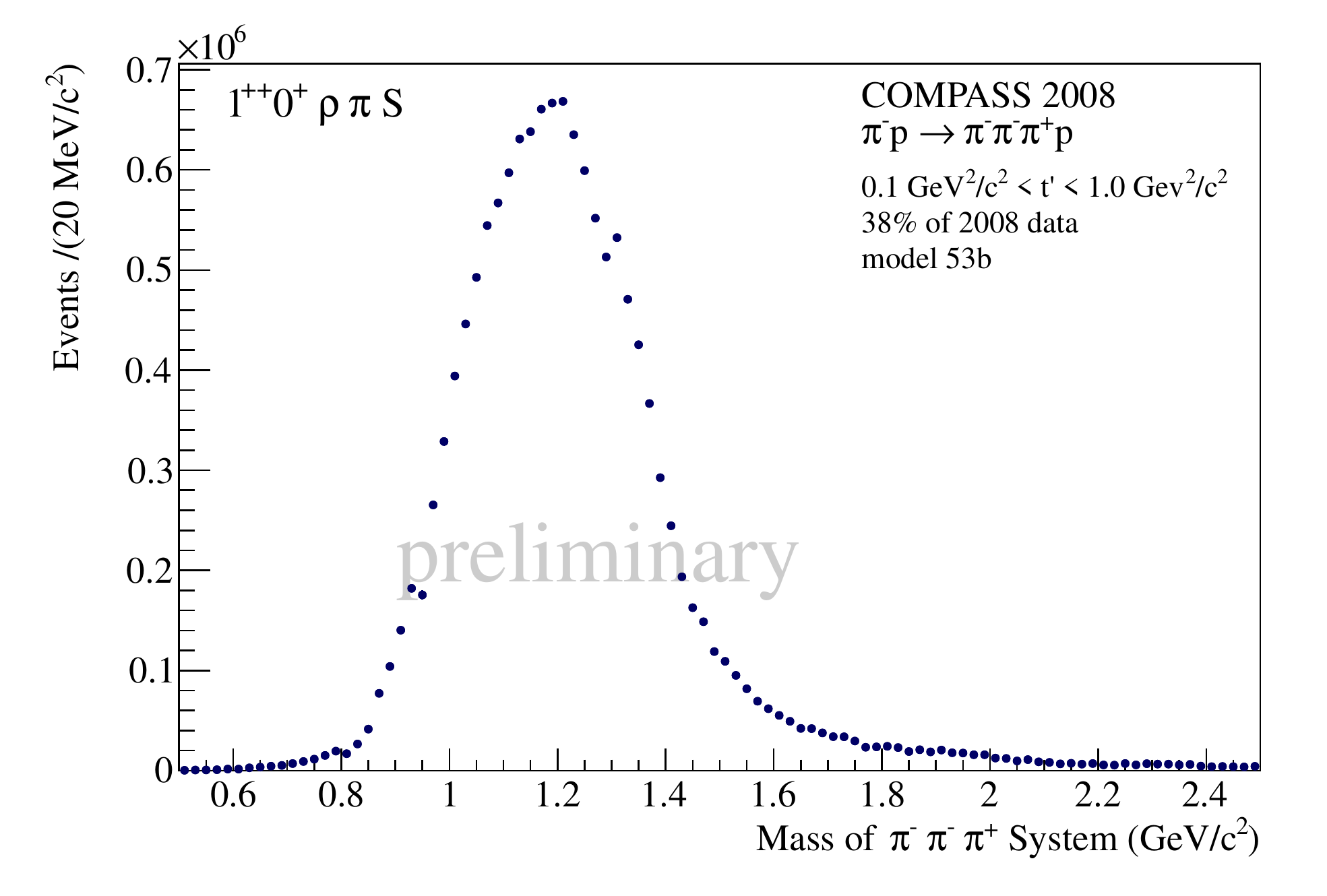}
}
%\qquad
\subfloat
[]
{
\label{pi2_2008}
\includegraphics[width=0.325\textwidth]{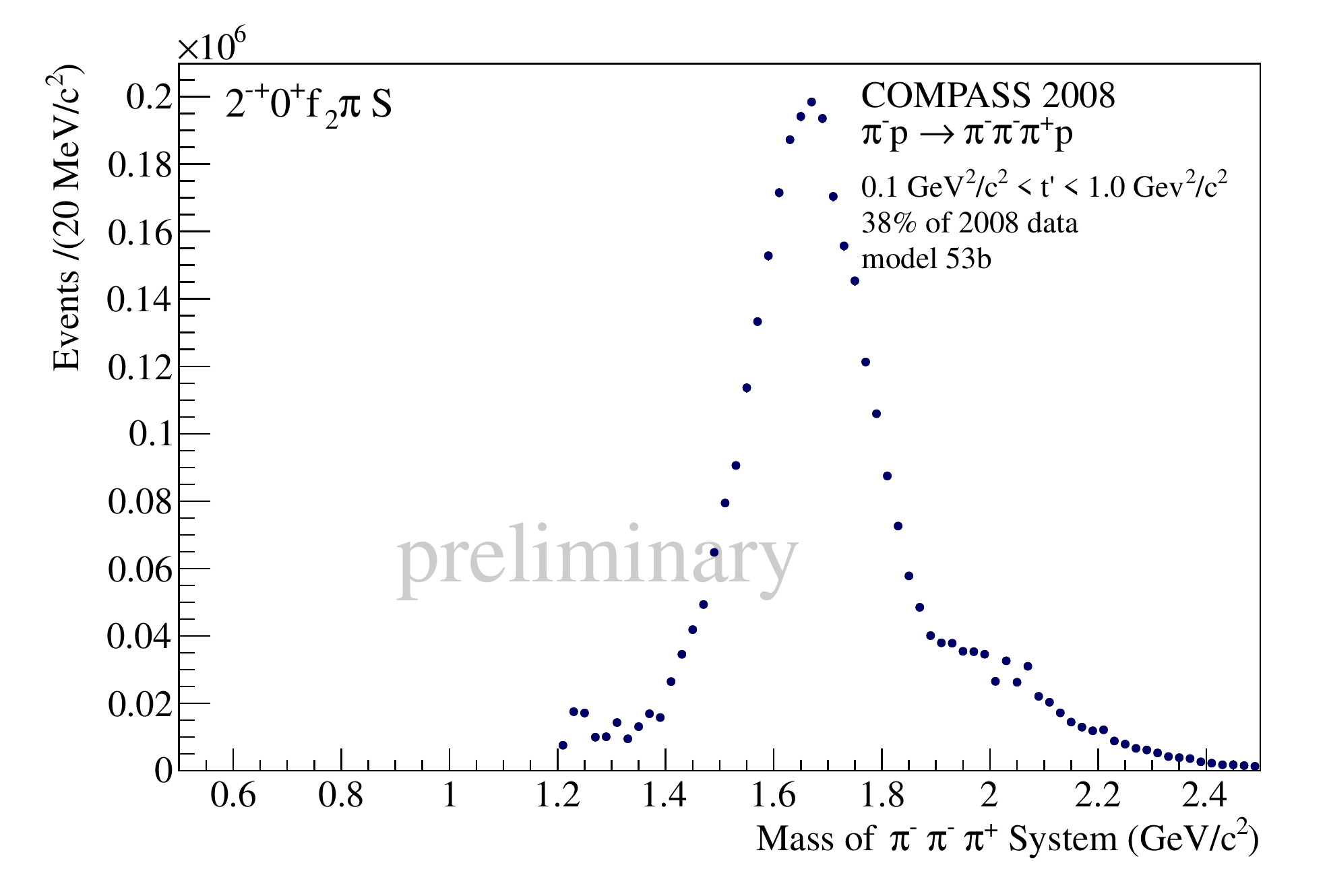}
}

\subfloat
[]
{
\label{pi1_2008}
\includegraphics[width=0.325\textwidth]{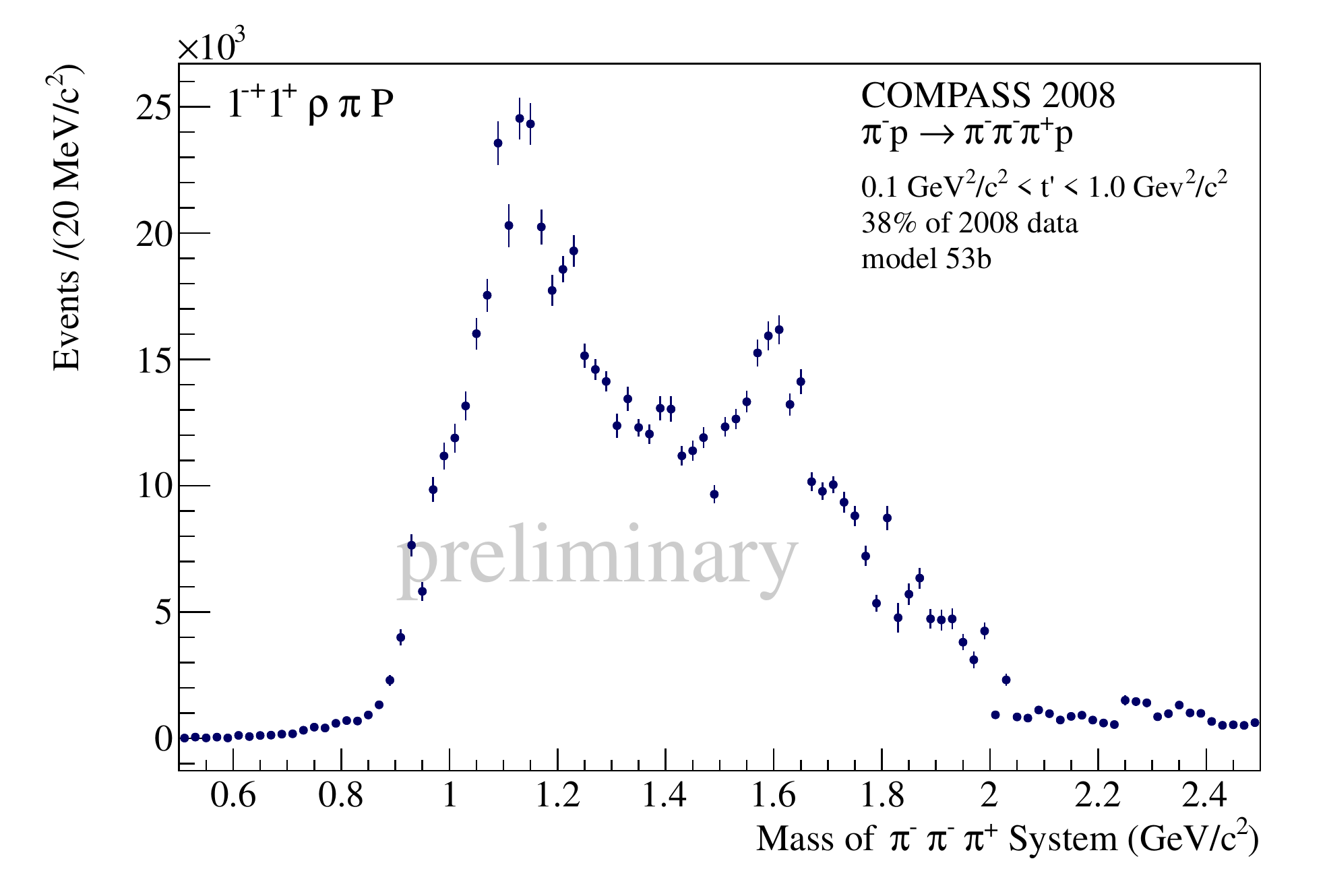}
}
%\qquad
\subfloat
[]
{
\label{phasea1_2008}
\includegraphics[width=0.325\textwidth]{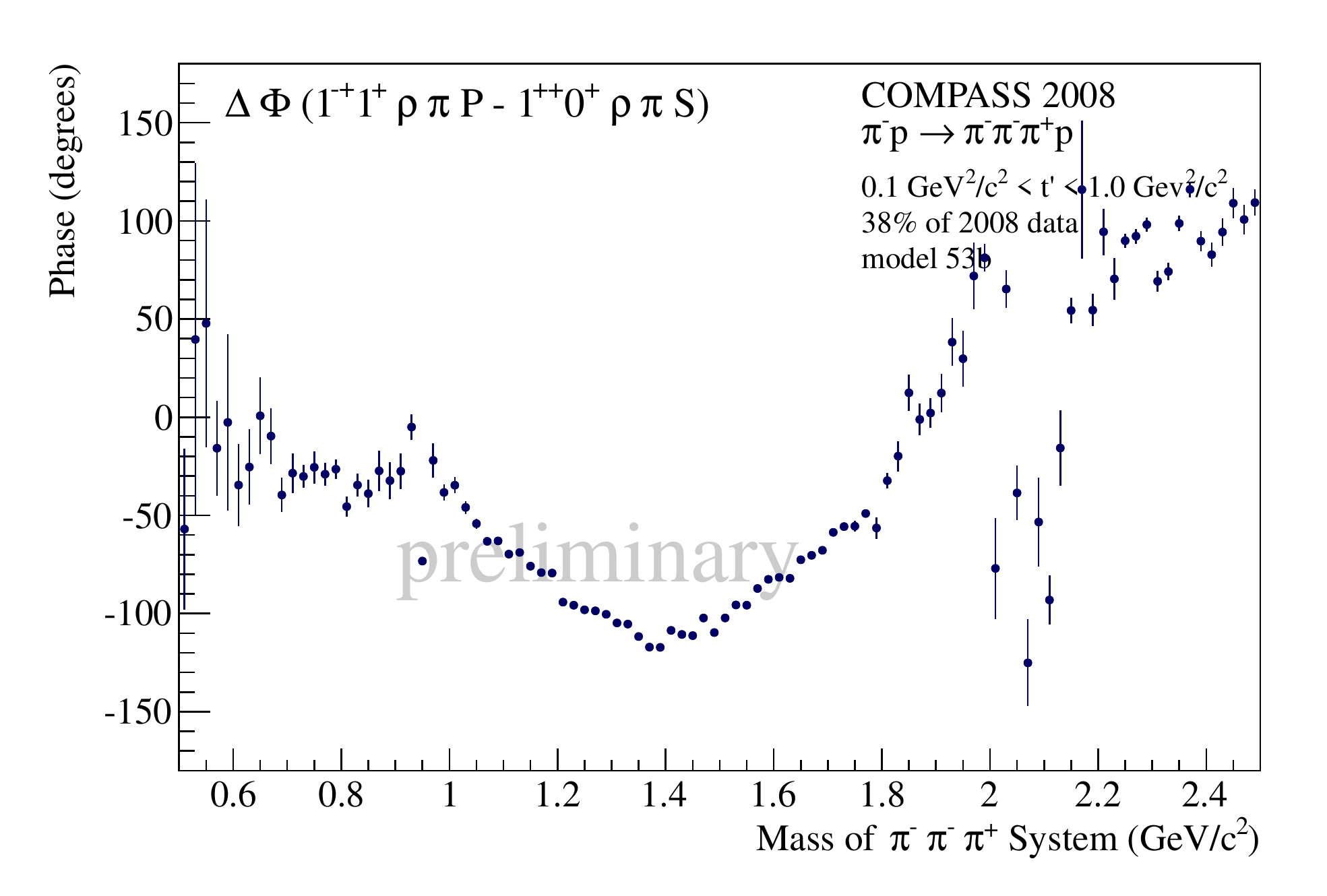}
}
%%\qquad
\subfloat
[]
{
\label{phasepi2_2008}
\includegraphics[width=0.325\textwidth]{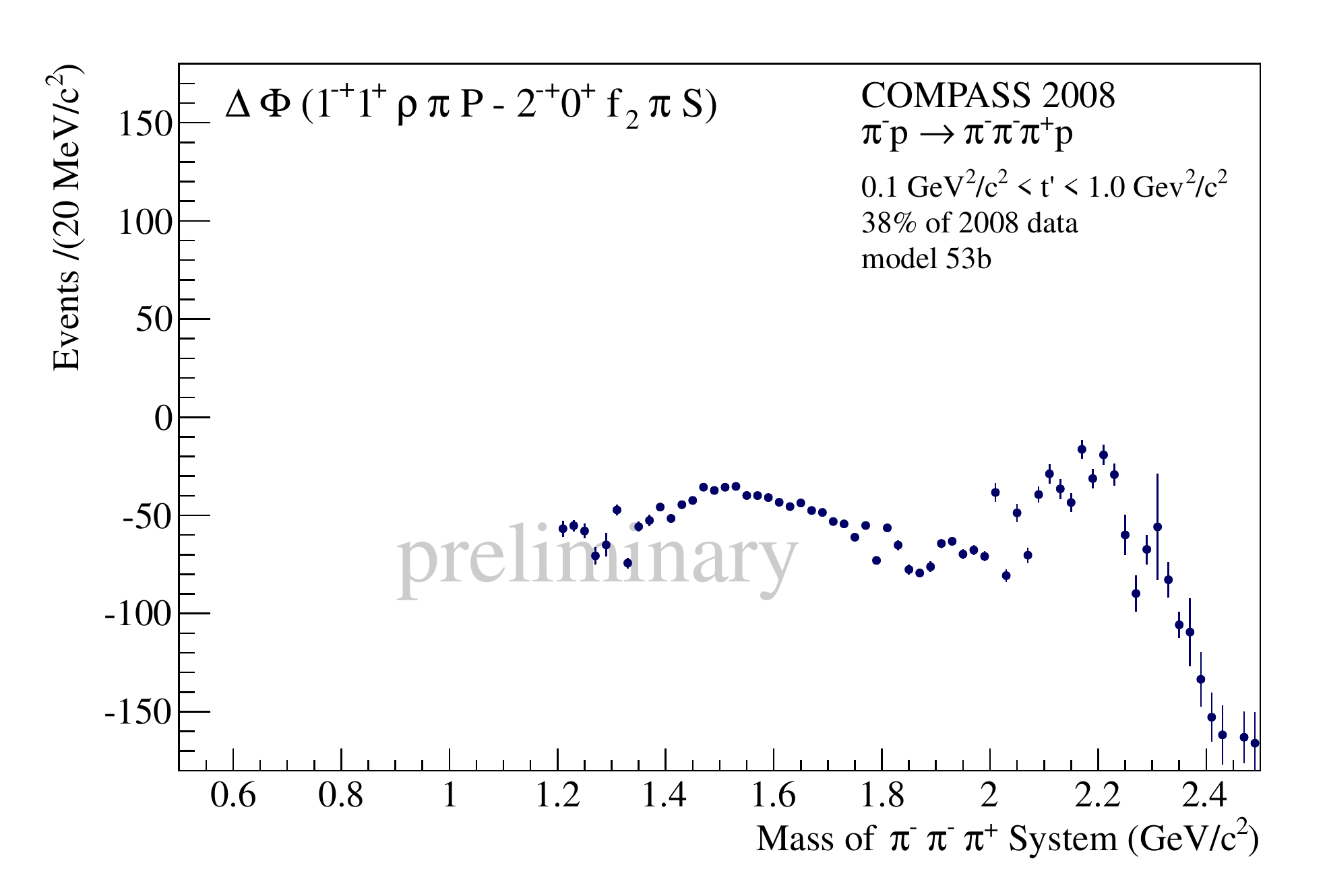}
}
\caption{Intensities of major waves: (a) $2^{++}1^+ \rho \pi D$, (b) $1^{++}0^+ \rho\pi S$ and (c) $2^{-+}0^+ f_2 \pi S$, no mass-dependent fit applied yet. (d) Intensity of the spin-exotic wave $1^{-+}1^+ \rho \pi P$, and phase differences of this wave with respect to the (e) $1^{++}0^+ \rho\pi S$, and the (f) $2^{-+}0^+ f_2 \pi S$ waves, no mass-dependent fit applied yet (H$_2$ target).}
\label{fitsum2008}
\end{figure}

\enlargethispage{\baselineskip}

\clearpage

 established states have been studied. Figure~\ref{phasea1}
shows the phase difference to the $1^{++}0^+\rho \pi S$
wave (Fig.~\ref{a1}), which clearly rises between 1.5 and 1.9 GeV/$c^2$. Figure~\ref{phasepi2} shows that the exotic wave is phase locked with the $\pi_2(1670)$ resonance in the $2^{-+}0^+ f_2 \pi S$ wave, which can be explained by the presence of a $\pi_1(1600)$  with mass and width similar to the $\pi_2(1670)$.
Further details of this analysis can be found in \cite{3pic}.\\
A similar analysis was applied to the large data set with the liquid hydrogen target. The waveset, used for the 2004 analysis, was extended due to 150 times larger statistics. Additional waves, with $J = 5$ and $J = 6$, $M = 2$ and waves described by a sixth isobar ($f_0(1500$) were implemented. No $t'$ dependence was introduced. A mass-dependent fit was not performed yet. The result of a fit in 20 MeV mass bins can be seen in Fig.~\ref{fitsum2008}.\\
Again the intensity of the three major waves is plotted (Fig.~\ref{fitsum2008}a-c), while Fig.~\ref{pi1_2008} shows the exotic wave, where in contrast to the previous results the signal-to-background ratio is much smaller. This effect can be observed in several $M=1$ waves, which is discussed in more detail in Sec.~\ref{M}. But nevertheless a peak centered at 1.65 GeV/$c^2$ and a respective phase motion (Fig.~\ref{phasea1_2008}) with respect to the $1^{++}0^+\rho \pi S$ wave (Fig.~\ref{a1_2008}) can be seen in the same mass region as in the 2004 analysis. Also the flat phase difference (Fig.~\ref{phasepi2_2008}) between the spin-exotic and the $2^{-+}0^+ f_2 \pi S$ wave (Fig.~\ref{pi2_2008}) is in good agreement with the fit results on a Pb target. The peak-like structure at 1.1 GeV/$c^2$ is not yet completely understood. The fact that no change of the phase difference can be seen in the corresponding
mass region in Fig.~\ref{phasea1_2008}, suggests a non-resonant mechanism. Further studies are underway.

\subsection{Dependence of $\mathbold {\pi^- \pi^- \pi^+}$ Production on the Target Material}\label{M}
A striking finding is the target material dependence of the population of the $M$ sub-states of the produced waves 
which becomes evident from the analysis of data taken with a Pb target (2004, 2009) and with the lH target (2008), and 
which is shown here for the 2008/2009 data sets. Due to different amounts of statistics both data samples are normalized to 
the integral of the narrow $2^{++}1^+ \rho \pi D$ wave in the region between 1.1 and 1.6 GeV/$c^2$. A clear difference in 
population of $M$ sub-states can be seen in comparison of the two data samples. As an example the total intensities
of $J^{PC} = 1^{++}$waves with different $M$ projections are shown in Fig.~\ref{Mfig}. The population of $M = 1$ waves is
 significantly higher for the heavy nuclear target (Fig.~\ref{M12009}) than for hydrogen (Fig.~\ref{M12008}). 
At the same time, the population of $M=0$ waves is reduced for Pb (Fig.~\ref{M02009}) as compared to 
hydrogen (Fig.~\ref{M02008}). With the applied normalization the total intensity for both projections, however, 
is rather similar (Fig.~\ref{Ma2008} and Fig.~\ref{Ma2009}). The origin of this dependence is yet unexplained.\\

\begin{figure}[h!]
\subfloat
[]
{
\includegraphics[width=0.33\textwidth]{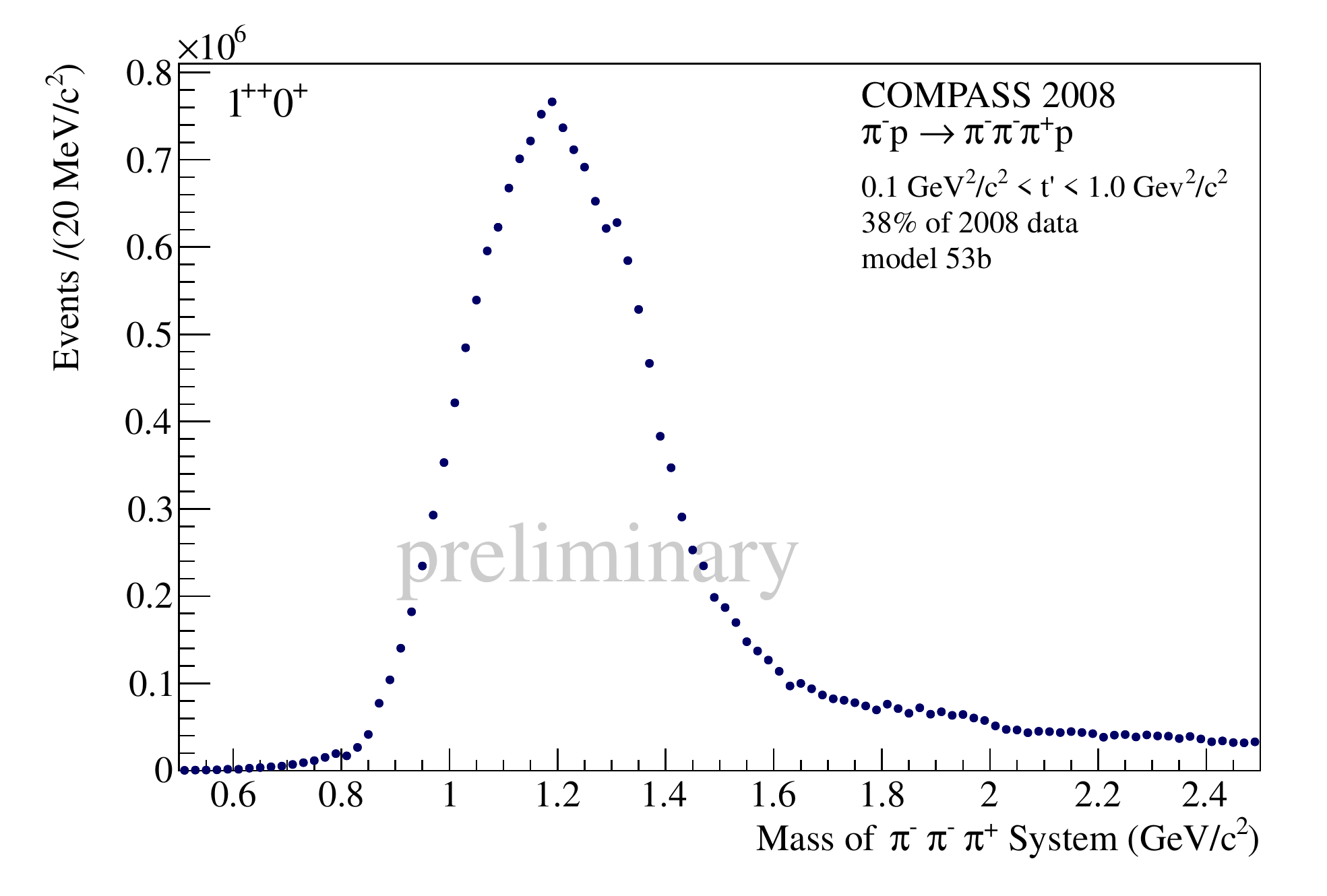}
\label{M02008}
}
%\quad
\subfloat
[]
{
\includegraphics[width=0.33\textwidth]{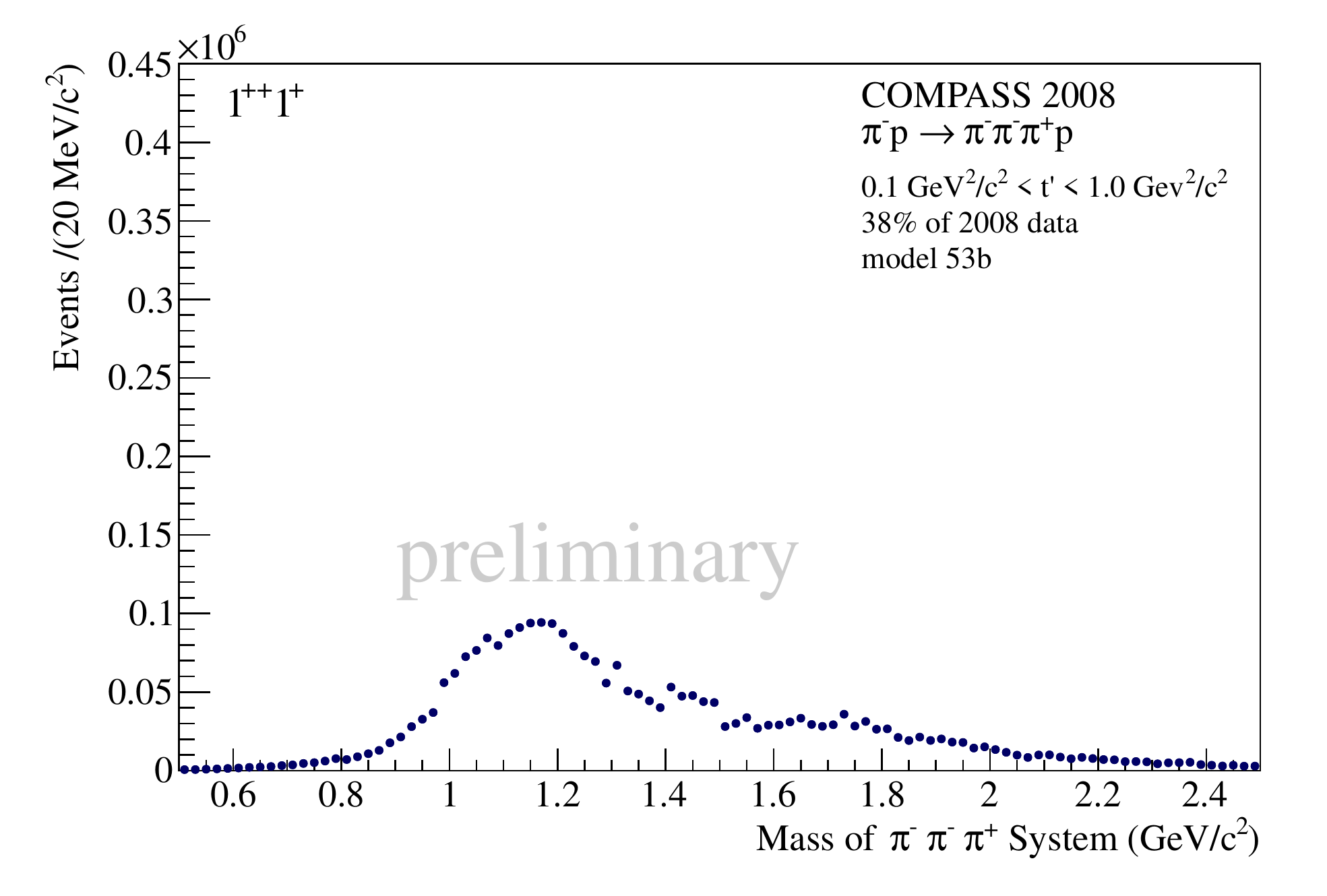}
\label{M12008}
}
%\quad
\subfloat
[]
{
\includegraphics[width=0.33\textwidth]{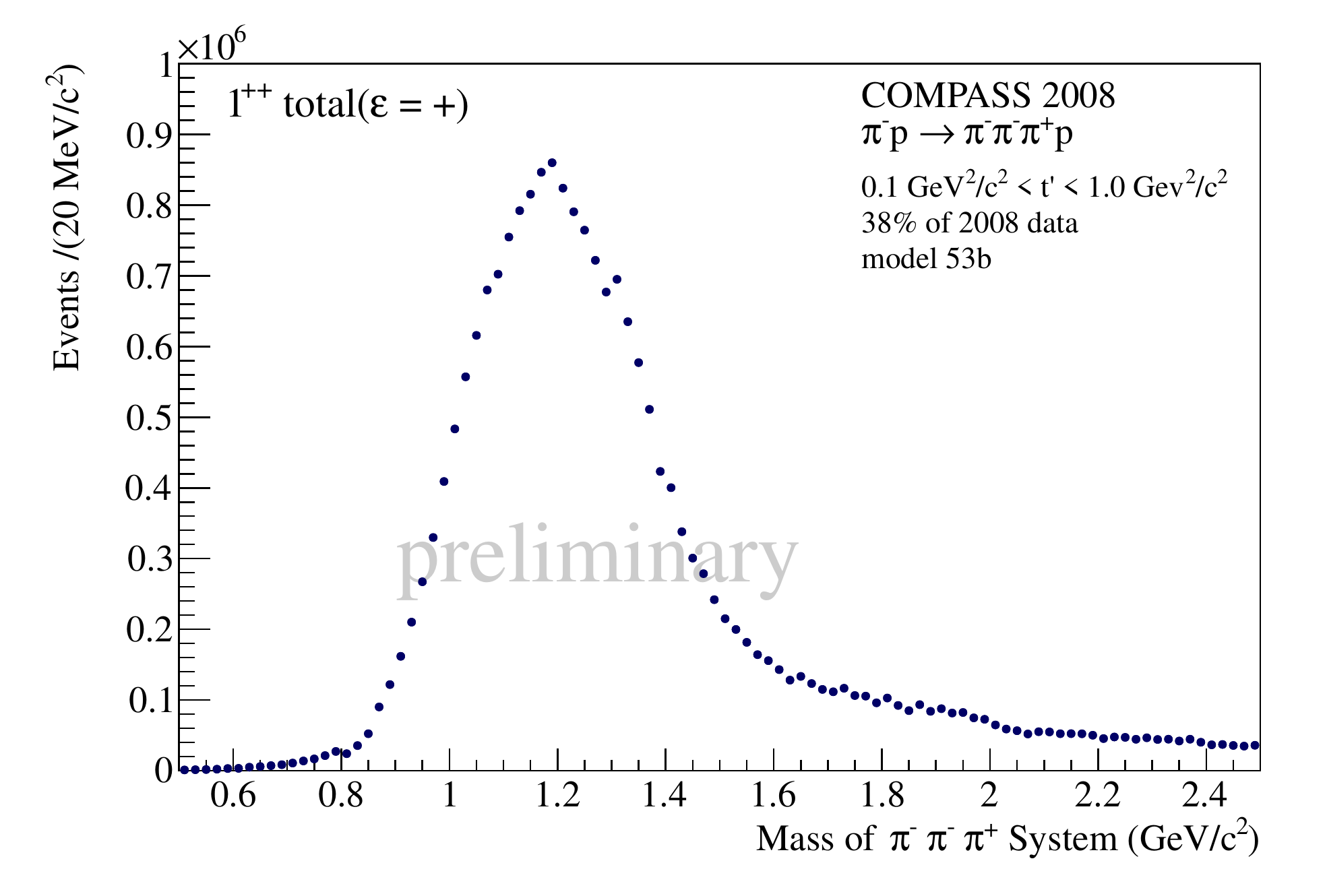}
\label{Ma2008}
}
%\caption{Intensity sum of  $J^{PC} = 1^{++}$ waves with $M=0$ (a), $M=1$ (b), and for both $M$ projections (c) (Pb target).}
%\label{M2004}
%\end{figure}

%\begin{figure}[h!]
\subfloat
[]
{
\includegraphics[width=0.33\textwidth]{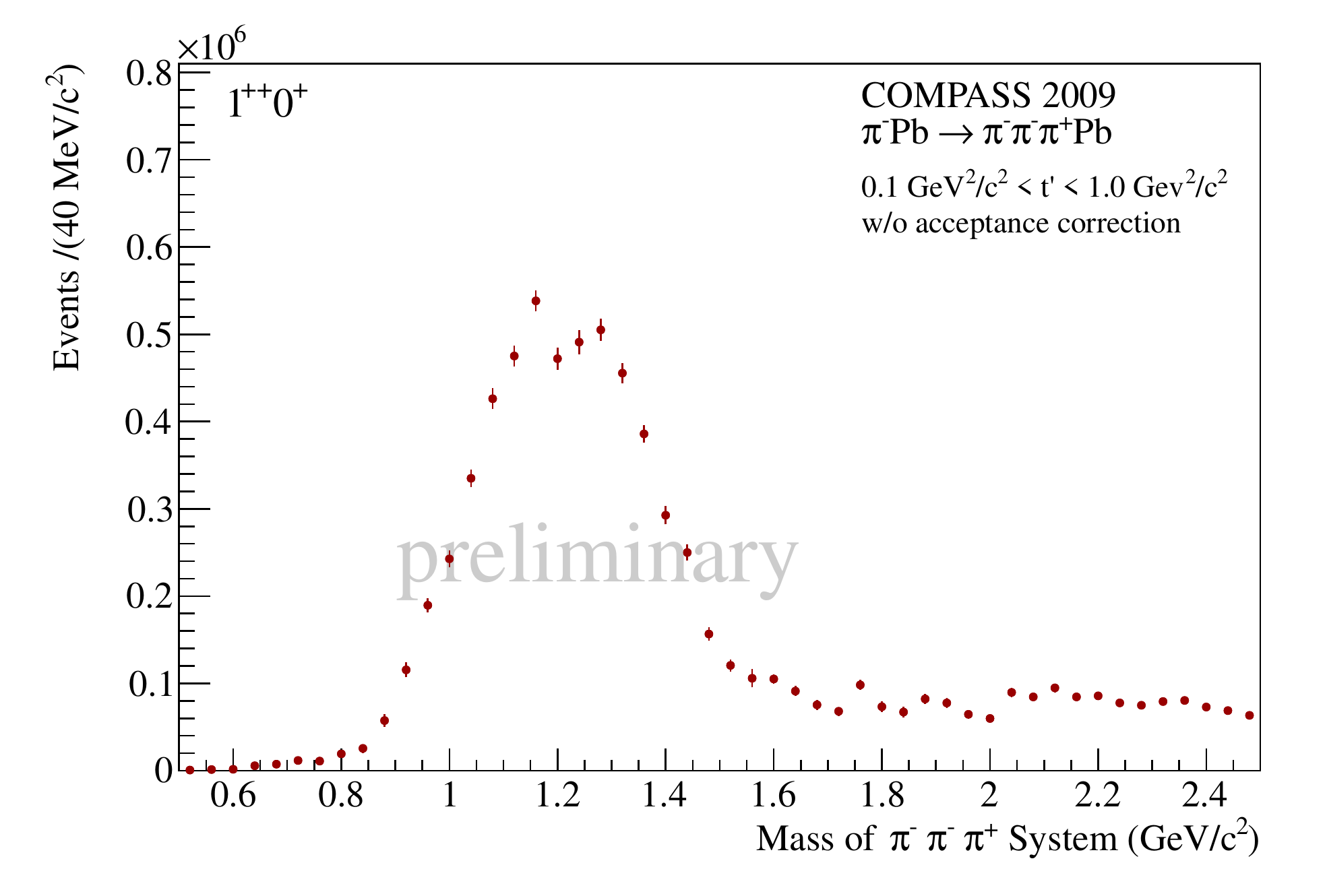}
\label{M02009}
}
%\quad
\subfloat
[]
{
\includegraphics[width=0.33\textwidth]{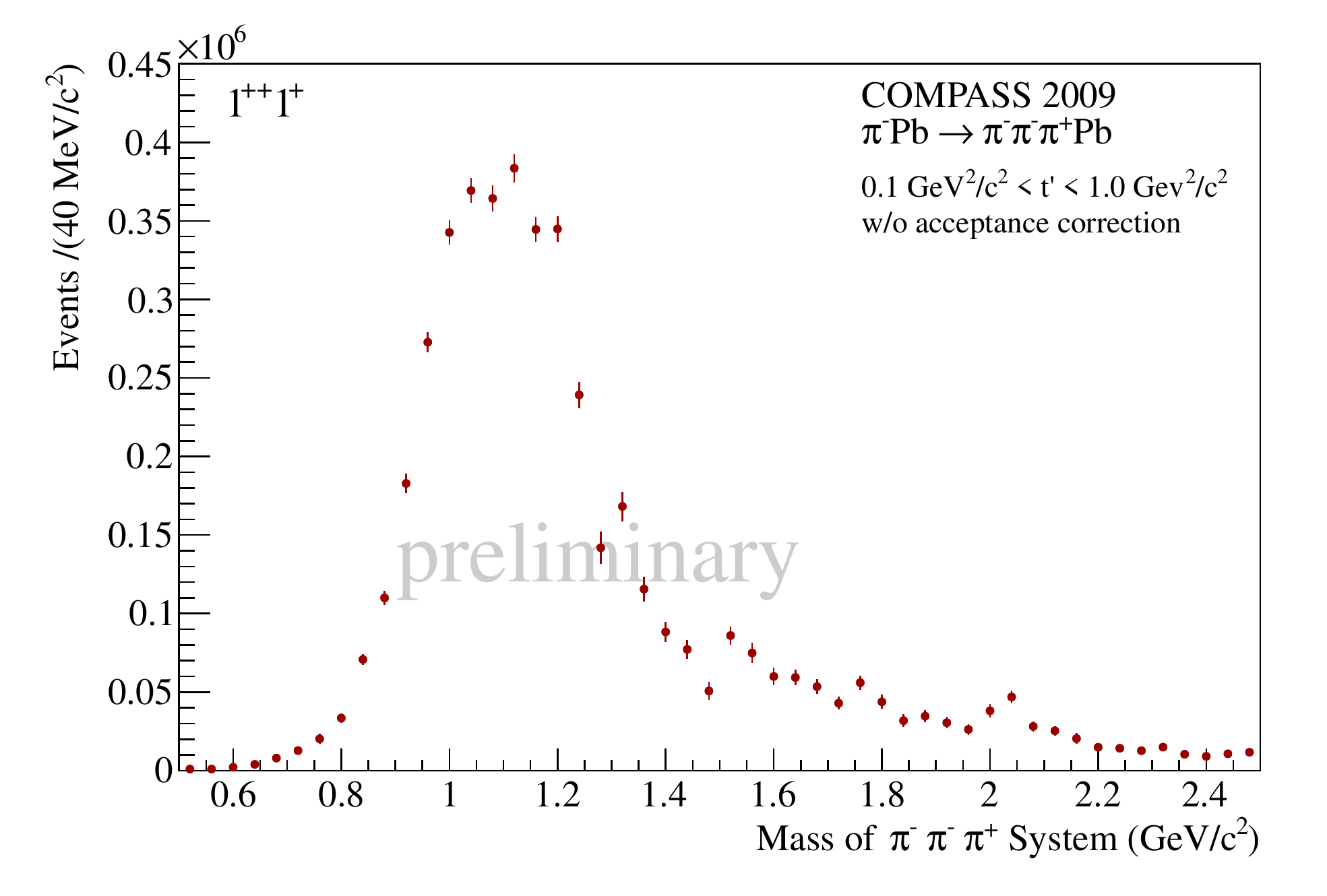}
\label{M12009}
}
%\quad
\subfloat
[]
{
\includegraphics[width=0.33\textwidth]{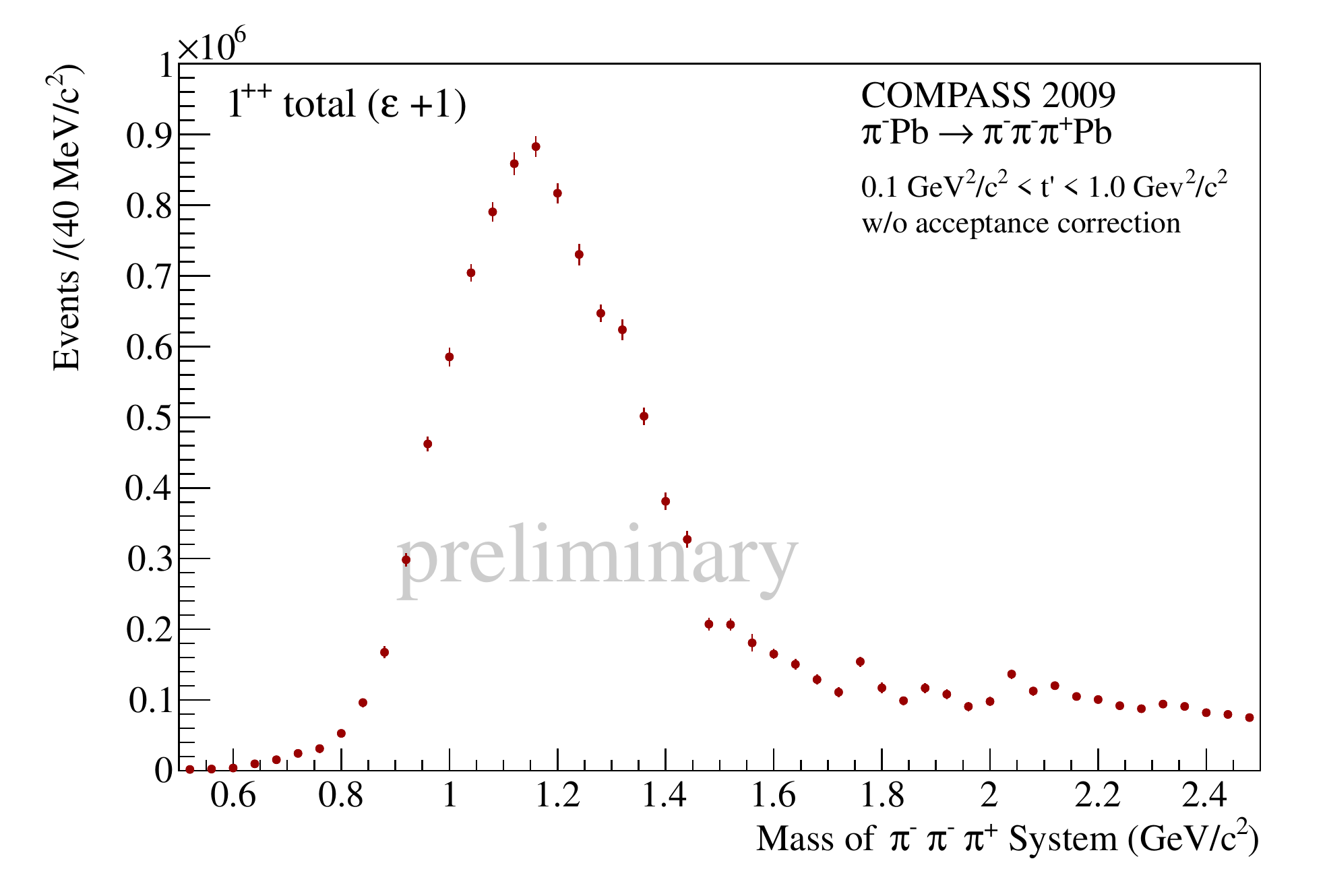}
\label{Ma2009}
}
\caption{Intensity sum of  $J^{PC} = 1^{++}$ waves with $M=0$: (a) H$_2$, (d) Pb, $M=1$: (b) H$_2$, (e) Pb, and for both $M$ projections: (c) H$_2$, (f) Pb, same scaling, Pb data normalized to H$_2$ data.}
\label{Mfig}
\end{figure}

\section{Conclusions}
The COMPASS spectrometer is a powerful tool to investigate the light meson spectrum. The 2004 pilot run data on a Pb target show a strong signal of the spin-exotic $\pi_1(1600)$, observed in the $1^{-+}1^+ \rho \pi P$ wave. \\
During the hadron spectroscopy data taking campaign in 2008 COMPASS recorded a large diffractive data sample on a proton target which in the $\pi^- \pi^- \pi^+$ final state exceeds the available world statistics by a factor of 10.
Preliminary fit results confirm the presence of the $\pi_1(1600)$ in good agreement with the 2004 analysis, also on the lH target.
A striking difference, however, is the dependence of the production strength for different $M$ sub-states of waves on the target material which indicates a suppression of $M=1$ states for light targets. Fits in mass and $t'$ bins are underway as well as leakage studies. Also Deck-like effects will be studied, in order to get a deeper insight into non-resonant contributions. This knowledge will lead to an improved mass-dependent fit to clarify the resonance nature of structures seen in the data, especially in the spin-exotic $J^{PC} = 1^{-+}$ wave.

%%%%%%%%%%%%%%%%%%%%%%%%%%%%%%%%%%%%%%%%%%%%%%%%%%%%%%%%%%%%%%%%%%%%%%%%%%%%%%%%%
% acknowledgements (optional)
\acknowledgements{%
 This work is supported by the
the German BMBF, the
Maier-Leibnitz-Labor der LMU und TU M\"unchen, and the DFG cluster of excellence
{\em Origin and Structure of the Universe}.
}

%\enlargethispage{3\baselineskip}
%%%%%%%%%%%%%%%%%%%%%%%%%%%%%%%%%%%%%%%%%%%%%%%%%%%%%%%%%%%%%%%%%%%%%%%%%%%%%%%%%
% bibliographic items can be constructed using the LaTeX format in SPIRES
% see http://www.slac.stanford.edu/spires/hep/latex.html
% SPIRES will also supply the CITATION line information; please include it

%
%%%%%%%%%%%%%%%%%%%%%%%%%%%%%%%%%%%%%%%%%%%%%%%%%%%%%%%%%%%%%%%%%%%%%%%%%%%%%%%%%

}  % do not remove

%%% Local Variables: 
%%% mode: latex
%%% TeX-master: "../hadron2011.tex"
%%% End: 

\end{document}